\documentclass[11pt]{article}
\hoffset = - 0.9 truecm \voffset=-1.4 truecm
\def\beq{\begin{equation}}   \def\eeq{\end{equation}}
\def\bea{\begin{eqnarray}}  \def\eea{\end{eqnarray}} \def\nn{\nonumber}
\def\noi{\noindent} \def\beeq{\begin{eqnarray}}
\def\eeeq{\end{eqnarray}}
\def\lsim{\raise0.3ex\hbox{$<$\kern-0.75em\raise-1.1ex\hbox{$\sim$}}}
\def\gsim{\raise0.3ex\hbox{$>$\kern-0.75em\raise-1.1ex\hbox{$\sim$}}}

\usepackage{epsfig}

\begin{document}

\begin{titlepage}

\begin{flushright}
LAPTH-918/02\\
LPT-Orsay 02-52\\
IPPP/02/31\\
DCPT/02/62\\
June 2002
\end{flushright}
\vspace{1.cm}

\begin{center}
\vbox to 1 truecm {}
{\large \bf A NLO calculation of the hadron-jet cross section in 
photoproduction reactions}
\par \vskip 3 truemm
\vskip 1 truecm {\bf M. Fontannaz$^{(a)}$, J. Ph. Guillet$^{(b)}$, G.
Heinrich$^{(c)}$} \vskip 3 truemm

{\it $^{(a)}$ Laboratoire de Physique Th\'eorique, UMR 8627 CNRS,\\
Universit\'e
Paris XI, B\^atiment 210, 91405 Orsay Cedex, France}\\

 \vskip 3 truemm

{\it $^{(b)}$ LAPTH, UMR 5108 du CNRS associ\'ee \`a l'Universit\'e
de Savoie, \\ BP 110, Chemin de Bellevue, 74941 Annecy-le-Vieux
Cedex, France}

\vskip 3 truemm

{\it $^{(c)}$ Department of Physics, University of Durham, Durham DH1 
3LE, England}

\vskip 2 truecm

\normalsize

\begin{abstract}

We study the photoproduction of large-$p_T$ charged hadrons in $e\,p$
collisions, both for the inclusive case and for the case
where a jet in the final state is also measured.
Our results are obtained by a NLO generator of
partonic events. We discuss the sensitivity 
of the cross section to the renormalisation and factorisation scales, 
and to various fragmentation function parametrisations. 
The possibility to constrain the parton densities in the proton 
and in the photon is assessed.
Comparisons are made with H1 data for inclusive charged hadron production.

\end{abstract}

\end{center}

\end{titlepage}

\baselineskip=20 pt
\section{Introduction}
\hspace*{\parindent} The photoproduction of large-$p_T$ jets, photons
or hadrons are privileged reactions to study QCD and to measure the
parton distributions in the proton and in the photon. In the past, the
interest was mainly focused on the production of jets
\cite{1r,2r}, and more recently on the
production of photons \cite{3r}. Particularly interesting are the dijet
cross sections \cite{4r,5r}, or the photon-jet cross sections \cite{6r},
because the measurement of two jets or particles in the final state 
allows one to constrain the incoming
parton kinematics and to explore the parton distributions in an accurate
way. \par

In this paper we present results concerning the large-$p_T$
photoproduction of a charged hadron and a jet, $e\,p\to h^{\pm}+jet$. 
This reaction offers several
interesting features in comparison with dijet or photon-jet reactions.
With respect to the latter, the hadron-jet cross section is much higher\,;
the observation of a hadron is also easier than that of a photon.
Compared to the dijet reaction, the hadron-jet one is also easier to
measure, this fact being particularly true at small $p_T$
($\sim$~5\,-10~GeV) where it is difficult to model the underlying event
contribution and to unambiguously measure the transverse momentum of two
jets. It should also be possible to explore a larger rapidity domain for
the hadron since there is no cone "hitting the edge" of the detector. \par

These features are important when one focuses on the measurement of the
parton distributions in the photon. As is well known, there are two
contributions to photoproduction cross sections\,: the so-called direct
and resolved contributions. The resolved contribution is important at
small values of $p_T$ and at large positive rapidities, a kinematic
domain that the hadron-jet reaction should allow one to explore. \par

The theoretical description of the hadron-jet cross sections requires the
knowledge of the fragmentation functions of quarks and gluons into
hadrons. This might appear as a disadvantage with respect to the
jet-jet reaction. However one has to note that hadronisation
corrections are also needed in jet reactions to describe the evolution
of partons into hadrons and they are not totally under control.
Moreover fragmentation functions are now well measured in LEP
experiments and several recent NLO parametrisations of quark and gluon 
fragmentation
functions are available \cite{7r,8r,9r}. These fragmentation 
functions have been tested in inclusive charged
hadron production in $p\bar{p}$, $\gamma p$ and $\gamma\gamma$
collisions by the authors of ref. \cite{8r}\,; they found a good
agreement between theory and data, which confirms that the fragmentation
functions are under control. \par

In this paper we concentrate on the hadron-jet physics and assess the
possibility to constrain the parton distribution functions
of the photon and of the proton.
In jet-jet reactions, it is
usual to constrain the momentum of the incoming partons by means of the
variables $x^{\rm{p},\gamma}_{obs} = (p_{T}^{jet_1} e^{\pm \eta^{jet_1}} + p_{T}^{jet_2}
e^{\pm \eta^{jet_2}})/2E^{\rm{p},\gamma}$ and of the variables 
$x^{\rm{p},\gamma}_{LL} = p_{T}^{jet_1} (e^{\pm \eta^{jet_1}}+e^{\pm \eta^{jet_2}})/2E^{\rm{p},\gamma}$,
where $E^{\rm{p},\gamma}$ are the energies of the
incoming proton respectively photon. (We follow the HERA convention with the
proton momentum oriented toward the positive $z$-axis and the photon
momentum toward the negative $z$-axis). For the Born contributions,
with only two jets in the final state, the variables $x^{\rm{p},\gamma}$ exactly
correspond to the longitudinal momentum fractions of the partons 
in the proton and in the
photon. When higher order QCD corrections are considered, the variables
$x^{\rm{p},\gamma}$ do not fix the initial momenta any more, but they still put
useful constraints on these momenta. The situation is different in the
hadron-jet case because the hadron momentum $p_T^h$ is only a fraction
of the momentum of the outgoing parton, such that $x^{\rm{p},\gamma}$ (with
$p_{T}^{jet_1}$ replaced by $p_T^h$) do not even for the Born cross-section
correspond to the incoming parton momenta anymore. Therefore it
is interesting to study the usefulness of such variables in the
hadron-jet case, and how they can help to measure the proton and photon
parton distributions. \par

The results presented in this paper are based on a NLO Monte Carlo
program which generates events containing 2 or 3 partons in the final
state. One of them fragments into a large-$p_T$ hadron, and the others
are recombined into jets. This NLO event generator provides a flexible
approach to implement experimental cuts and to calculate a large
variety of observables. \par

The theoretical framework and the Monte Carlo program 
are presented in section 2. In section 3 we compare 
theoretical predictions for the inclusive (i.e. no
jet observed) cross section  with H1 data~\cite{18r}. 
The emphasis is put on the study of the sensitivity of the cross section to 
the factorisation and renormalisation scales, and to different 
fragmentation function  parametrisations. 
In section 4 we study the hadron-jet cross section and explore the 
distributions $d\sigma/dx^{\rm{p},\gamma}$, in particular their sensitivity to the gluon densities in the 
proton and the photon. 
Section 5 is the conclusion.

\section{Theoretical framework}
\hspace*{\parindent}
The NLO Monte Carlo program used in this paper has already been described in
refs. \cite{10r,11r} in which the photoproduction of isolated prompt
photons is studied. Therefore we only give a few indications on
the general structure of the program here and discuss the new features
specific to the photoproduction of hadrons. \par

In photoproduction events, the electron acts like a source of
quasi-real photons whose spectrum can be described by the
Weizs\"acker-Williams formula

\beq \label{1e} f_{\gamma}^e(y) = {\alpha_{em} \over 2 \pi} \left \{ {1
+ (1 - y)^2 \over y} \ \ln {Q^2_{max}(1-y) \over m_e^2 y^2} - {2(1-y)
\over y} \right \} \ . \eeq

\noi The quasi-real photon then either takes part {\it directly} in the
hard scattering process, or it acts as a composite object, being a
source of partons which take part in the hard subprocess. The latter
mechanism is referred to as {\it resolved} process and is parametrised
by the photon structure functions $F_{a/\gamma}(x_{\gamma},Q^2)$. Thus
the distribution of partons in the electron is a convolution

\beq \label{2e} F_{a/e}(x_e,M) = \int_0^1 dy\ dx_{\gamma}\
f_{\gamma}^e(y)\ F_{a/\gamma}(x_{\gamma},M) \delta (x_{\gamma}y - x_e)
\eeq

\noi where in the ``direct'' case $F_{a/\gamma}(x_{\gamma},M) =
\delta_{a\gamma} \delta(1- x_{\gamma})$. \par

The production of the final hadron $h$ with momentum $P_h$ is described
by a fragmentation function $D_a^h(z, M_F)$ where $z$ is the 
fraction of the longitudinal momentum of the parton $a$
carried away by the hadron $h$. The production cross section of a
large-$p_T$ hadron and a jet is written as a convolution of the
distributions of initial partons, the fragmentation of the final parton
and the hard scattering cross sections

\bea \label{3e} &&d\sigma^{ep\to h\, jet}(P_p, P_e, P_h, P_{jet}) =
\sum_{a,b,c} \int dx_e \int dx_p \int dz \ F_{a/e}(x_e,M) \
F_{b/p}(x_p,M)\nn \\ &&\times d\widehat{\sigma}^{ab\to
c\, jet} (x_pP_p, x_eP_e, {P_h/z}, P_{jet}, \mu, M, M_F)
D_c^h (z,M_F) \ .\eea

\noi The hard cross sections $d\widehat{\sigma}^{ab\to c\, jet}$ are
calculated at the NLO accuracy. They are expansions in powers of
$\alpha_s(\mu )$

\beq \label{4e} d\widehat{\sigma}^{\gamma b\to c\, jet} =
\alpha_s(\mu) \ d\widehat{\sigma}^{\gamma b\to c\ jet}_{BORN} +
\alpha_s^2(\mu) \ d\widehat{\sigma}^{\gamma b\to c\, jet}_{HO}(\mu , M, M_F) +{\cal
O}(\alpha_s^3)\eeq

\beq \label{5e} d\widehat{\sigma}^{a b\to c\, jet} = \alpha_s^2(\mu)\
d\widehat{\sigma}^{ab\to c\, jet}_{BORN} + \alpha_s^3(\mu)\
d\widehat{\sigma}^{ab\to c\, jet}_{HO}(\mu , M, M_F) +{\cal
O}(\alpha_s^4)\ . \eeq

\noi In expressions (\ref{3e}), (\ref{4e}) and (\ref{5e}), we have
explicitly written the dependence on the large scales $\mu$, $M$ and
$M_F$. (For simplicity we choose the same factorisation scale $M$ 
for the incoming photon and proton.) 
The cross section (\ref{3e}) would not depend on these large scales 
if it were calculated to all orders in $\alpha_s(\mu )$. But after
truncation of the series (\ref{4e}) and (\ref{5e}), the cross section
will depend on $\mu$, $M$ and $M_F$. The $\mu$\,-, $M$- and
$M_F$\,-\,dependence of the HO terms partially  compensates
the scale-dependence of the Born cross sections,
and in the next sections we shall study the sensitivity of $d\sigma^{ep
\to h \, jet}$ to the renormalisation scale $\mu$ and the
factorisation scales $M$ and $M_F$. \par

Let us now discuss the various components of formula (\ref{3e}). The
fragmentation functions $D_a^h(z, M_F)$ have been accurately measured
in LEP experiments and several NLO parametrisations of the latter are
now available \cite{7r,8r,9r}. There exist non-negligible differences
between the parametrisations of individual $D_a^h(z,M_F)$, especially
at large $z \ \gsim\ 0.8$, but for cross sections involving sums over the
flavours $a$ and over the hadrons $h$ the differences are tiny. In
photoproduction, the weights of the different flavours which contribute
to the cross section are identical to those of the
$e^+e^-$-annihilation reaction (at least for the direct contribution).
Therefore the differences between the individual contributions should
be smoothed when summed over to form the cross section. In the next section
we shall check the sensitivity of the photoproduction cross section to
the fragmentation functions. \par

We now turn to another important component of formula (\ref{3e}), the
quark and gluon distributions in the photon. In this paper we use a new
NLO parametrisation of these distributions, the AFG02 parametrisation
\cite{12r}. The AFG02 parametrisation is an evolution of the AFG
\cite{13r} parametrisation. The new distributions are more flexible\,:
for instance the shape and normalisation of the non-perturbative gluon
distribution can be modified. This fact allows one to study the
sensitivity of cross sections to the gluon distribution. The
normalisation of the non-perturbative quark distributions can also be
adjusted. Contrarily to the AFG distributions, the AFG02 distributions
also contain the bottom quark distribution. In this paper we shall
 use the default AFG02 parametrisation which is almost identical
to the AFG parametrisation\,; the only differences are the number of
flavours and the value of $\Lambda_{\overline{MS}}^{(4)} = 300$~MeV
($\Lambda_{\overline{MS}}^{(4)} = 200$~MeV was used in the AFG
parametrisation in agreement with the value of
$\Lambda_{\overline{MS}}^{(4)}$ determined ten years ago). As a result,
due to a faster QCD evolution, the AFG02 distributions are slightly
higher at small $x$ and lower at large $x$ than the AFG
distributions. \\
For the parton distributions in the proton, we use 
MRST99 (g\,$\uparrow$)~\cite{mrst99} as default. 
For the hadron-jet cross section, we will  also use  
the new MRST01~\cite{mrst01} and CTEQ6M~\cite{cteq6} sets for comparison. 

We use a strong coupling constant  calculated  
by solving exactly (i.e. without expansion in log\,${Q^2/\Lambda^2}$) 
the two-loop renormalisation group equation. 
We work with $N_f=4$ active flavours 
as default. Using $N_f=5$ the total cross section increases by about 5\%.  
 
Expression (\ref{3e}) is calculated via a MC code which generates $2
\to 2$ and $2 \to 3$ parton configurations according to weights given
by the subprocess cross sections and the distribution functions. A
phase space slicing method is used to isolate and to analytically
calculate the soft and collinear singular contributions of the $2
\to 3$ cross sections. The soft divergences are cancelled by the
corresponding divergences contained in the virtual corrections to the
$2 \to 2$ processes (UV divergences are removed by renormalisation in
the $\overline{MS}$ scheme). The remaining collinear singularities are
factored out and absorbed in the distribution and fragmentation functions 
using the $\overline{MS}$ scheme. \par

This Monte Carlo code, which uses the event generator Bases/Spring \cite{14r},
is a partonic event generator\,: it contains negative weights coming for
instance from the virtual corrections to the Born cross sections. It is
very flexible and allows one to study various cross sections involving
a large-$p_T$ hadron and jets. Experimental cuts are easily taken into
account, as well as different jet algorithms. In this paper, we use the
$k_T$-algorithm \cite{15r} to define the jets.

\section{Large-p$_{\bf T}$ hadron inclusive cross section}
\hspace*{\parindent}
In this section we study the photoproduction of large-$p_T$ inclusive
hadrons. Similar studies and comparisons with data have already been
performed in several publications \cite{8r,16r,17r}. Therefore we shall not
carry out an exhaustive study of this reaction, but we shall concentrate
on the sensitivity of the cross section to the factorisation and
renormalisation scales, and to different parametrisations of the
fragmentation functions.\par

\begin{figure}[htb]
\begin{center}
\mbox{\epsfig{file=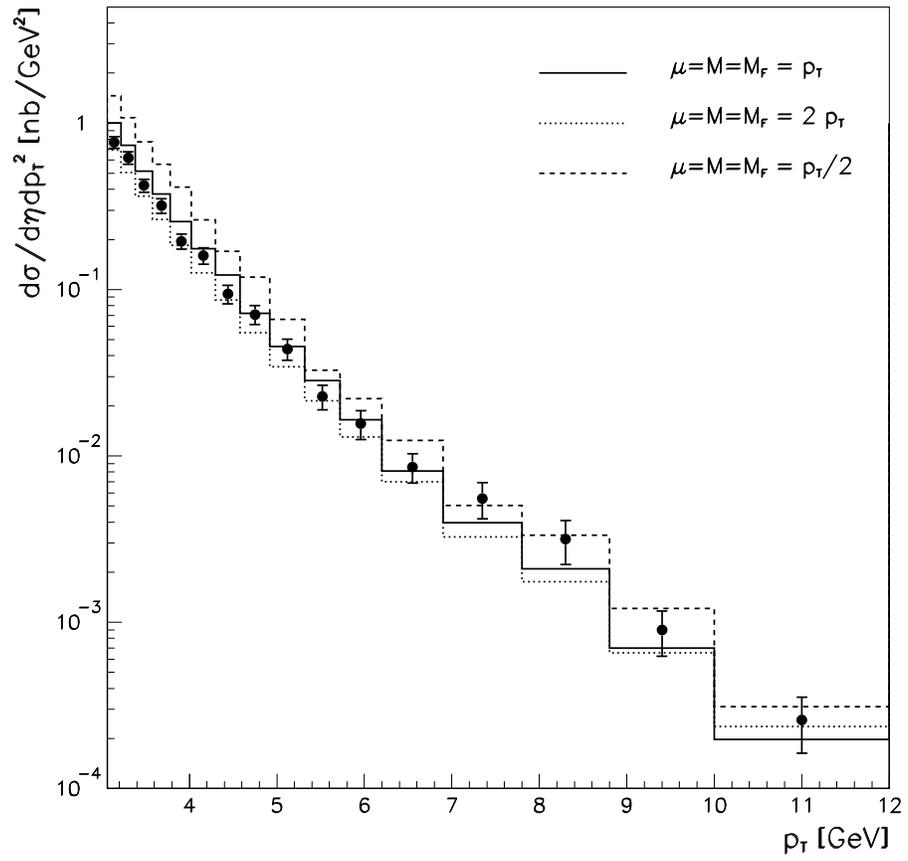,height=13cm}}
\end{center}
\caption{$d\sigma/d\eta\,dp_T^2$ for 3 different scale choices compared to H1 data.}
\label{fig1}
\end{figure}

In Fig.~\ref{fig1}, we display a study of the scale dependence. The theoretical
curves are compared to H1 data \cite{18r} corresponding to the
following kinematic conditions. The parameters of the
Weizs\"acker-Williams formula are $Q_{max}^2 = 0.01$\,GeV$^2$ and $0.3
\leq y \leq 0.7$. The $e\,p$ center of mass energy is $\sqrt{S_{ep}} =
300$~GeV. The cross section for large-$p_T$ charged hadrons is measured
in the pseudo-rapidity domain $|\eta| \leq 1$. The
theoretical curves are obtained with the BFGW fragmentation functions~\cite{9r}
and we use $\Lambda_{\overline{MS}}^{(4)} = 300$\,MeV. \\
\noi We can see from Fig.~\ref{fig1} that the data are described fairly well, in
particular that no intrinsic $k_T$ is needed to describe the data at low $p_T$,
contrarily to what has been found in the E706 experiment on fixed target 
inclusive $\pi^0$ production~\cite{e706,lapi}.

\begin{figure}[htb]
\begin{center}
\mbox{\epsfig{file=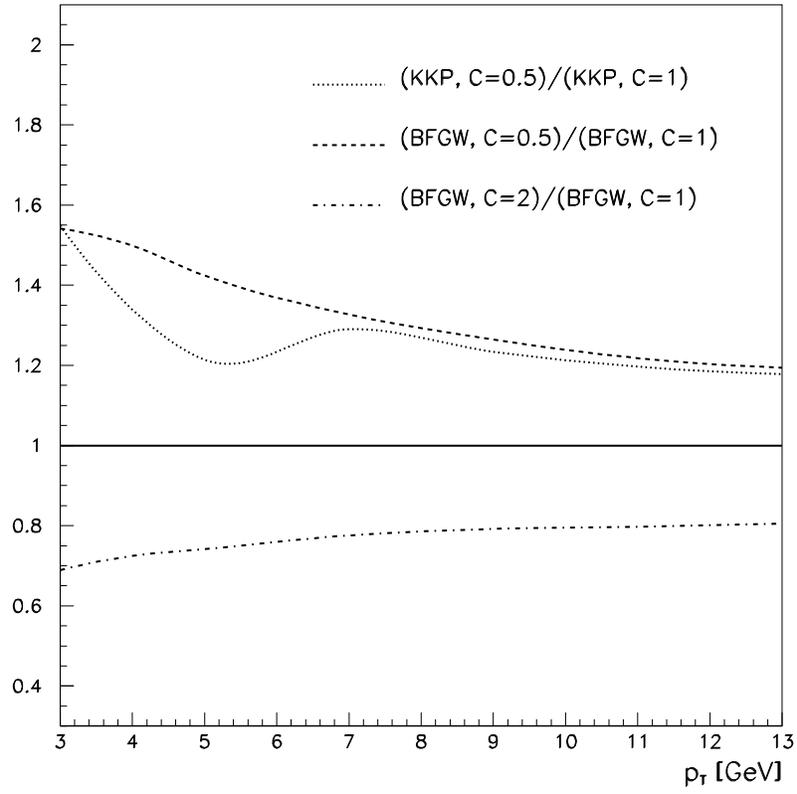,height=12cm}}
\end{center}
\caption{$d\sigma/dp_T$ calculated with scales 
$\mu=M=M_F= C\,p_T$ ($C$=0.5,\,2) normalised to $d\sigma/dp_T$ at 
$\mu=M=M_F= p_T$.}
\label{fig2}
\end{figure}

In Fig.~\ref{fig2} we show on a
linear scale the ratios of the cross sections calculated with $M = M_F = \mu =
C\,p_T$ ($C=0.5,2$) to the cross section calculated at $C=1$, where 
$p_T$ is the transverse momentum of the final state hadron. 
Results obtained with the KKP~\cite{8r} fragmentation function parametrisations
are also displayed\footnote{ 
The dip at $p_T\sim 5$\,GeV in the KKP ratio may stem from the 
fact that the charm threshold is at $M_F=2m_c$ in these 
paramtrisations. Therefore there is no charm contribution below 
$p_T\sim 6$\,GeV in the case $C=0.5$, whereas for $C=1$ the charm contributes 
already at 3\,GeV. In the BFGW parametrisations the charm threshold 
is at $M_F=m_c$, such that the charm contributes at $p_T\geq 3$\,GeV
even for $C=0.5$.}. 
We clearly see a strong dependence of the cross section on the scales for
$p_T \ \lsim \ 7$\,GeV. Below 7\,GeV the perturbative calculation is not
reliable, the HO corrections being very large.
Moreover, for the choice $C=1/2$ we explore a range of the factorisation scales
($M=M_F=p_T/2\sim 1.5$\,GeV) which is very far from the kinematic region 
where the fragmentation functions have been constrained.

\clearpage

We also find a strong scale dependence of the rapidity distribution
$d\sigma/d\eta$ which is measured in the range 3\,GeV $\leq p_T \leq 
12$\,GeV. In Table~\ref{muM} we give the results of a study of 
$\Delta\sigma = \int_0^{.5} d\eta {d\sigma \over d\eta}$ in which we
separately vary $\mu$ and $M$, keeping fixed $M_F = p_T$. The upper
number is the total cross section (direct + resolved) and the lower
number is the ratio $r = HO/(Born + HO)$. The variation of the cross
section with $C_{\mu}=\mu/p_T$ is very strong for small values of $C_{\mu}$
($C_{\mu} \sim 0.5$) corresponding to large values of $\alpha_s(\mu)$.
There is no
region where the cross section $\sigma (C_M, C_{\mu})$ is almost
independent of $C_M$ and $C_{\mu}$, and the ratio $r$ is always large. 
We again conclude that the
theoretical predictions are not reliable for $p_T \sim 3$~GeV, a
$p_T$-region which gives an important contribution to the cross section 
integrated over 3\,GeV $\leq p_T \leq 12$\,GeV. 

\begin{table}[h]
\begin{center}
\begin{tabular}{|l||c|c|c|c|}
\hline
&&&&\\
&$C_M$&0.5&1&2\\
&&&&\\
\hline
\hline
&&&&\\
$C_{\mu}$=0.5&${\bf \sigma}$&{\bf 3.941}&{\bf 4.509}&{\bf 4.807}\\
&$r$&{\it (0.54)}&{\it (0.42)}&{\it (0.31)}\\
&&&&\\
\hline
&&&&\\
$C_{\mu}$=1&${\bf \sigma}$&{\bf 2.580}&{\bf 3.079}&{\bf 3.430}\\
&$r$&{\it (0.57)}&{\it (0.48)}&{\it (0.41)}\\
&&&&\\
\hline
&&&&\\
$C_{\mu}$=2&${\bf \sigma}$&{\bf 1.855}&{\bf 2.264}&{\bf 2.564}\\
&$r$&{\it (0.58)}&{\it (0.52)}&{\it (0.46)}\\
&&&&\\
\hline
\end{tabular}
\end{center}
\caption{$\mu,M$-dependence of the total cross section integrated 
over $0\le\eta\le 0.5$ and 3\,GeV$< p_T <$12\,GeV. The fragmentation scale has been kept
fixed to $M_F=p_T$. The bold numbers denote the total cross section in nb, the 
numbers in italic represent the ratio $r=HO/(Born+HO)$. One can see that there 
is no region of stability for a value of $p_{T}^{\rm{min}}$ as low as 3\,GeV.}
\label{muM} 
\end{table}
The situation improves when we study the sensitivity to scale changes 
for larger values of $p_T$. For instance at $p_T=7$\,GeV -- although the 
sensitivity is still large when we vary all scales by the same factor $C$ 
(see Fig.~\ref{fig2}) --
we find a region where the cross section as a function of $M$ and $\mu$
has a flat behaviour. 
We can see from Table~\ref{muMpt7}, where $M_F$ is kept fixed to $p_T/2$,
that there is a stability region for $0.3\lsim \,C_{\mu}\,\lsim 0.5$ and 
$0.5\lsim \,C_{M}\,\lsim 2$. 
Fixing $\mu=0.4 \,p_T$ and $M=1.5 \,p_T$ at the saddle point of the surface given in
Table~\ref{muMpt7}, we also studied the dependence on $M_F$ and found that the
cross section varies by $\pm 2\%$ when $M_F$ is varied between $0.3\,p_T$ and
$p_T$. 
Therefore the scale sensitivity appears to be under control for 
large values of $p_T \,(p_T\,\gsim \,7$\,GeV). 
We shall pursue this study in the next
section for the hadron-jet cross section.  

\begin{table}[h]
\begin{center}
\begin{tabular}{|l||c|c|c|c|c|}
\hline
&&&&&\\
&$C_{M}$&0.3&0.5&1&2\\
&&&&&\\
\hline
\hline
&&&&&\\
$C_{\mu}$=0.3&${\bf \sigma}$&{\bf 0.254}&{\bf 0.243}&{\bf 0.229}&{\bf 0.224}\\
&$r$&{\it (0.25)}&{\it (0.06)}&{\it (-0.19)}&{\it (-0.40)}\\
&&&&&\\
\hline
&&&&&\\
$C_{\mu}$=0.5&${\bf \sigma}$&{\bf 0.220}&{\bf 0.222}&{\bf 0.223}&{\bf 0.228}\\
&$r$&{\it (0.38)}&{\it (0.27)}&{\it (0.13)}&{\it (0.03)}\\
&&&&&\\
\hline
&&&&&\\
$C_{\mu}$=1&${\bf \sigma}$&{\bf 0.176}&{\bf 0.184}&{\bf 0.192}&{\bf 0.200}\\
&$r$&{\it (0.45)}&{\it (0.38)}&{\it (0.29)}&{\it (0.23)}\\
&&&&&\\
\hline
\end{tabular}
\end{center}
\caption{$\mu,M$-dependence of the total cross section integrated 
over $|\eta|\leq 1$ at $p_T =7$\,GeV. The fragmentation scale has been
fixed to $M_F=p_T/2$. The bold numbers denote the total cross section in nb, the 
numbers in italic represent the ratio $r=HO/(Born+HO)$. In this case there 
is a rather flat region for small values of $\mu$ and large values of $M$, 
and the overall variation of the cross section due to scale changes is much smaller.}
\label{muMpt7} 
\end{table}

\begin{figure}[htb]
\begin{center}
\mbox{\epsfig{file=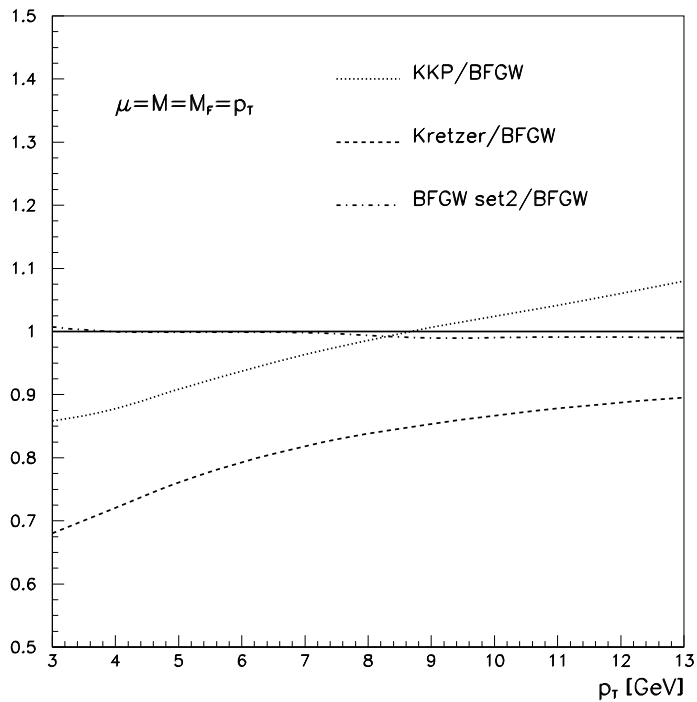,height=10.5cm}}
\end{center}
\caption{$d\sigma/dp_T$ with different fragmentation functions 
normalised to $d\sigma/dp_T$ with BFGW (set1) fragmentation functions, 
at the scales $\mu=M=M_F=p_T$ and the hadron rapidity integrated 
over the range $|\eta|<1$.}
\label{figffs}
\end{figure}

\begin{figure}[htb]
\begin{center}
\mbox{\epsfig{file=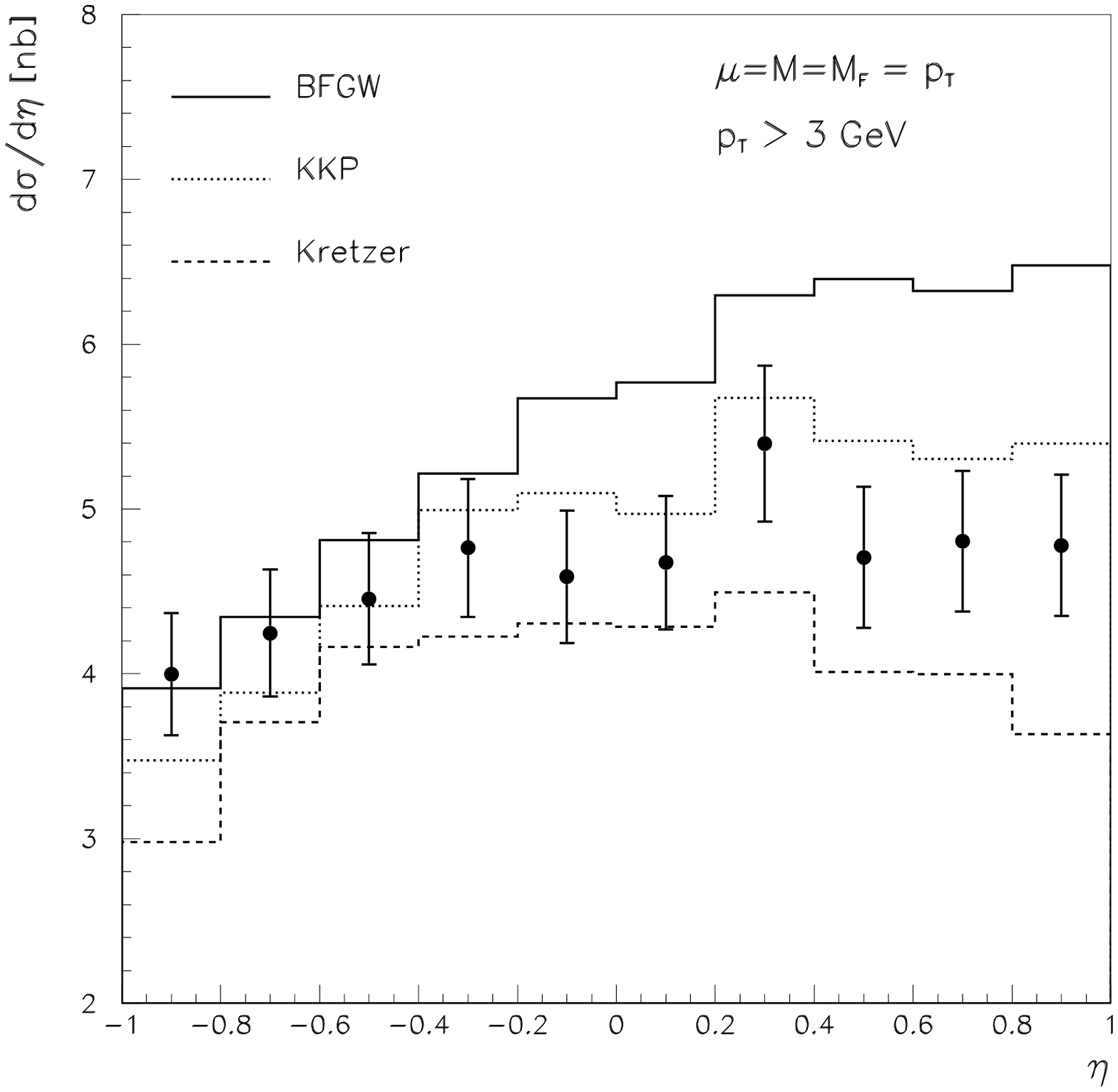,height=12.5cm}}
\end{center}
\caption{$d\sigma/d\eta$ with different sets of fragmentation 
functions compared to H1 data.}
\label{figrap}
\end{figure}
\begin{figure}[htb]
\begin{center}
\mbox{\epsfig{file=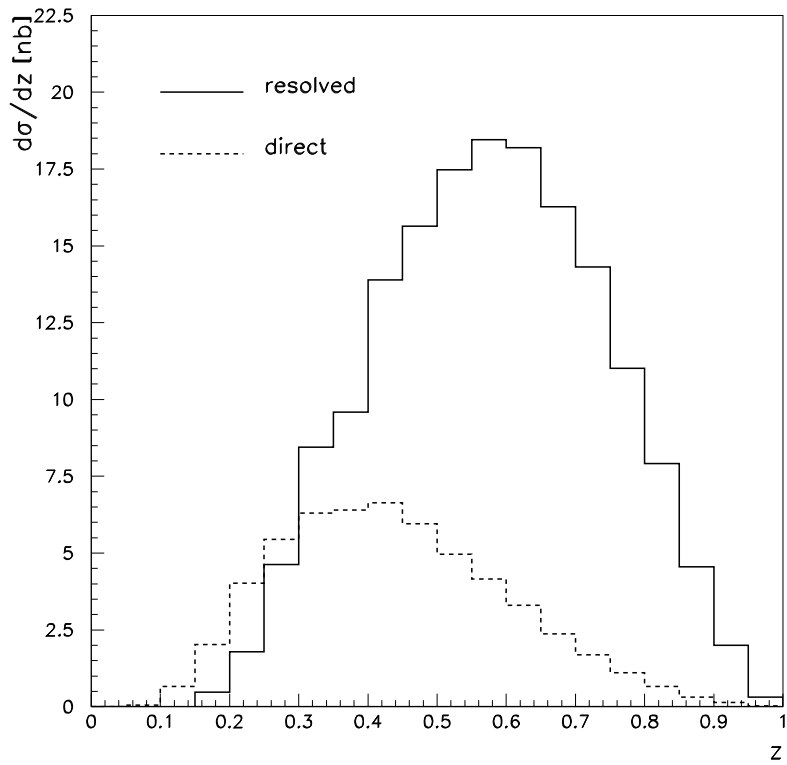,height=11.5cm}}
\end{center}
\caption{Distribution of the hadron momentum fraction $z$ for the direct and 
resolved contributions. (3\,GeV$\leq p_T\leq 12$\,GeV, 
$|\eta|<1$, $\mu=M=M_F=p_T$.)}
\label{figz}
\end{figure}

The sensitivity to the fragmentation function parametrisations~\cite{7r,8r,9r}
is less pronounced than the one due to scale variations. 
This result is illustrated in Fig.~\ref{figffs}. 
For $p_T>$ 7\,GeV, there is at most a 20\% difference between
the various parametrisations, the BFGW and KKP parametrisations being quite
close to each other. The BFGW set2 parametrisation has a slightly higher gluon
than the default BFGW for $0.2\,\lsim \,z\,\lsim \,0.6$ and a lower gluon for 
$z\,\gsim \,0.6$, due to a higher $N_g$ in the input parameter set. 
Since the HERA kinematics do not probe high $z$ values,  
the difference to the BFGW default set is negligible in our case, as can be seen 
from Fig.~\ref{figffs}. 
However, the sensitivity to the fagmentation functions is non-negligible for 
$p_T\sim$ 3\,GeV, a range explored by H1. In Fig.~\ref{figrap} we display 
a comparison between H1 data and theoretical results obtained with different 
sets of fragmentation functions. 
The dispersion of the results is quite 
large. In this regard it is instructive to
look at the distribution
$d\sigma/dz$, which is not a physical quantity (the momentum of the parton
"parent" of the hadron $h$ cannot be measured), but which gives interesting
indications on the average value $\langle z\rangle$ in the cross section, 
and on the variance $\langle z^2\rangle-\langle z\rangle^2$. 
The $z$-distributions are
displayed in Fig.~\ref{figz} for the direct and resolved components. 
One can see that two different ranges in $z$ of the fragmentation functions are
probed by the direct respectively resolved component. 
The mean value $\langle z\rangle$ is larger for the resolved
contribution\footnote{This is due to the fact that the cross section
for the production of partons 
behaves like $d\sigma/dp_T^{parton}\sim A/(p_T^{parton})^n\;(n>0)$, 
such that, using $p_T^{parton}=p_T^{h}/z$, the behaviour 
of the cross section is $d\sigma/dz\sim z^{n-2}$ convoluted with the 
fragmentation functions which decrease with increasing $z$. 
The value of $n$ is larger for the resolved part, 
leading to a larger value of $\langle z\rangle$.}. 
Since the various fragmentation function parametrisations 
differ mostly at high $z$, 
a change of parametrisation will have a different effect on the 
direct and on the resolved contribution. This effect can be seen in  
Fig.~\ref{figrap} where   
we observe that the differences between the various
parametrisations are more pronounced for positive rapidities, a region
corresponding to a large resolved contribution. We again verify in  
Fig.~\ref{figrap} that the value of $p_T^{\rm{min}}$=3\,GeV is too small 
to allow a reliable prediction, the dispersion of the results 
for the various parametrisations being too large at low $p_T$.

We further investigated the impact of using an expansion in log\,${Q^2/\Lambda^2}$
for $\alpha_s$ instead of the numerical solution of the two-loop 
renormalisation group equation, and found the result being about 10\% lower
when using the expansion in log\,${Q^2/\Lambda^2}$. 
We also compared our result to the one of Kniehl, Kramer and P\"otter~\cite{8r}.  
Using the KKP fragmentation functions and the expansion in log\,${Q^2/\Lambda^2}$
for $\alpha_s$, we obtain the result given in~\cite{8r} within the numerical 
errors.

\clearpage

\section{Hadron-jet cross section}

In this section we explore the features of the hadron-jet cross section. 
The input parameters are the same as for the inclusive cross section, and we
consider the cross section $d\sigma/d\eta^h\,d\eta^{jet}$ (where $\eta^h$ is the 
pseudo-rapidity of the observed charged hadron), integrated in the range
7\,GeV $\leq p_T^h\leq$ 15\,GeV and $E_T^{jet}>$ 5\,GeV. 
When there are two jets in the final state, we consider the jet of 
higher $E_T$. The jets are defined by the $k_T$-algorithm with the 
Snowmass merging rule. 

We shall also study the cross sections 
$d\sigma/dx^{\rm{p},\gamma}$, the variables 
$x^{\rm{p},\gamma}$ being either the $x_{obs}$ variables
\beq
x_{obs}^{\rm{p},\gamma}=\frac{p_T^h\,e^{\pm \eta^h}+E_T^{jet}\,e^{\pm
\eta^{jet}}}{2E^{\rm{p},\gamma}}
\label{xobs}
\eeq
or the 
$x_{LL}$ variables
\beq
x_{LL}^{\rm{p},\gamma}=\frac{p_T^h\,(e^{\pm \eta^h}+e^{\pm
\eta^{jet}})}{2E^{\rm{p},\gamma}}\;.
\label{xll}
\eeq
$E^{\rm{p},\gamma}$ are the energies of the incoming 
proton respectively photon and the plus sign in $e^{\pm \eta}$ 
corresponds to $x^{\rm{p}}$, the minus sign to $x^{\gamma}$. 
The main difference between $x_{obs}$ and $x_{LL}$ 
consists in the fact that the definition of $x_{LL}$ does not require
the measurement of the jet transverse energy. 
Further it has to be noted that -- contrarily to the dijet 
cross section -- the variables defined in (\ref{xobs}) and (\ref{xll}) 
do not even for the Born contribution coincide with the variables 
$x^{\gamma}_{parton}$ and $x^{\rm p}_{parton}$ in the parton densities of the photon and
 the proton. The latter are (for the Born contribution) given by
\beq
x^{\rm{p},\gamma}_{parton}=\frac{p_T^h}{z}\,
\frac{(e^{\pm \eta^h}+e^{\pm\eta^{jet}})}{2E^{\rm{p},\gamma}}\;.
\label{xtrue}
\eeq  
Therefore the partonic variables $x^{\rm{p},\gamma}_{parton}$ are larger 
by a factor $1/z$ than the variables $x^{\rm{p},\gamma}_{LL}$, 
and larger by a factor 
$(e^{\pm \eta^h}+e^{\pm\eta^{jet}})/(z\,e^{\pm \eta^h}+e^{\pm\eta^{jet}})$ 
than the variables $x^{\rm{p},\gamma}_{obs}$. At NLO, where a third parton in the final
state is involved, the situation is more complicated. 

\noi The variables $x_{obs}^{\rm{p},\gamma}$ and $x_{LL}^{\rm{p},\gamma}$ defined in (\ref{xobs}) and (\ref{xll}) 
differ by about 
$x_{obs}^{\rm{p},\gamma}/x_{LL}^{\rm{p},\gamma}\sim (z+1)/2z$ at central rapidities. 
This difference will be discussed below in the context of 
Figs.~\ref{figxga} and \ref{figxp}.

\begin{figure}[htb]
\begin{center}
\mbox{\epsfig{file=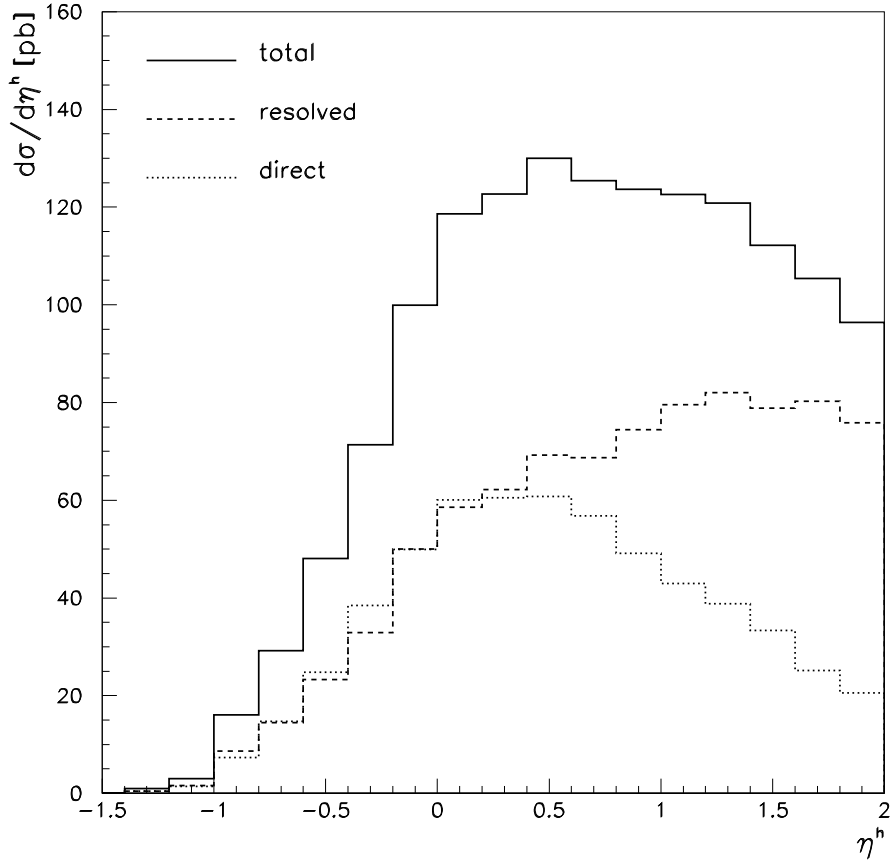,height=13cm}}
\end{center}
\caption{Rapidity distribution for the hadron-jet cross section 
$d\sigma/d\eta^h$ at the scales $\mu=M=M_F=p_T$, integrated over 
7\,GeV$\leq p_T\leq $15\,GeV, $E_T^{jet}\geq$ 5\,GeV, $|\eta^{jet}|\leq 2$.}
\label{fighjrap}
\end{figure}

Fig.~\ref{fighjrap} displays the direct and resolved contributions, calculated
with the scales $\mu=M=M_F=p_T$, as a function of $\eta^h$, integrated over 
$|\eta^{jet}|\leq 2$. The two contributions are comparable, except in the
forward region where the resolved one is much larger. However, we must keep in
mind that the pattern of the separate contributions to the cross section
depends on the choice of the scales $\mu,M$ and $M_F$\,; only the sum of the
resolved and direct contributions has a physical meaning and can be compared to
data. 

\begin{figure}[htb]
\begin{center}
\mbox{\epsfig{file=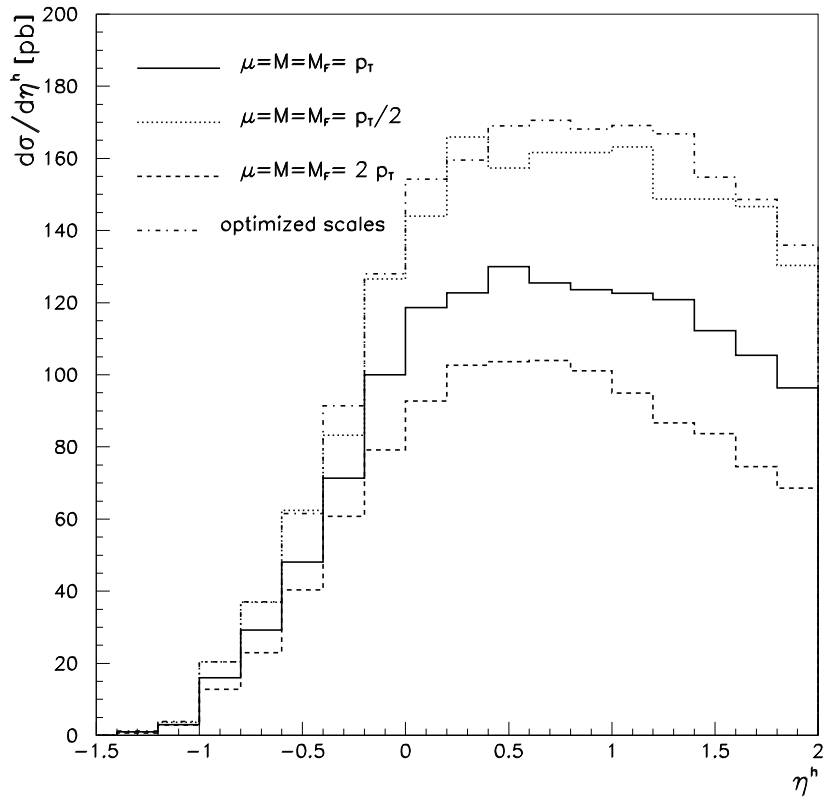,height=12cm}}
\end{center}
\caption{Scale dependence of $d\sigma/d\eta^h$ ($|\eta^{jet}|\leq 2$). }
\label{figrapsca}
\end{figure}

In Fig.~\ref{figrapsca} we display the scale dependence of the cross section 
$d\sigma/d\eta^h$ by varying all scales simultaneously, $\mu=M=M_F=C\,p_T$, 
with the parameter $C$ in the range $1/2\leq C\leq 2$. We see variations which
are similar to those observed in Fig.~\ref{fig2} in the inclusive case 
for $p_T\sim  7$\,GeV, where the cross section varies between +30\% and -20\% 
for $1/2\leq C\leq 2$. 
However, varying all three scales in the same way is a very rough measure of 
the scale dependence. A better study has been given in Table~\ref{muMpt7}
where we showed that the cross section exhibits a plateau in the scales 
$\mu$ and $M$ for  $0.3\lsim \,C_{\mu}\,\lsim 0.5$ and 
$0.5\lsim \,C_{M}\,\lsim 2$. We performed a similar investigation for the
hadron-jet cross section with the kinematic conditions of Fig.~\ref{figrapsca}.
However, contrarily to the preceding study, we look for an optimum of the
cross section in all three scales $\mu,M$ and $M_F$.  
That is, we look for optimal scales
where $d\sigma/d\mu=d\sigma/dM=d\sigma/dM_F=0$ \cite{stevenson}. 
We performed this
complete study for $d\sigma/d\eta^h$ with $\eta^h$ in the range $0.5\leq\eta^h\leq 1$.
The results are displayed in Table~\ref{muMhjet}. 

\begin{table}[h]
\begin{center}
\begin{tabular}{|c|l|c||c|}
\hline
&&&\\
$C_{M_F}$&$C_{\mu}^{\rm{opt}}$&$C_M^{\rm{opt}}$&$d\sigma/d\eta^h$\,[nb]\\
&&&\\
\hline
\hline
&&&\\
1&$\sim 0.21$&$\sim 1$&0.182\\
0.5&$\sim 0.3$&$\sim 1.5$&0.173\\
0.3&$\sim 0.3$&$\sim 1.5$&0.180\\
&&&\\
\hline
\end{tabular}
\end{center}
\caption{Scale optimisation for  the hadron-jet cross section integrated 
over $0.5\leq\eta^h\leq 1$, $|\eta^{jet}|\leq 2$, 7\,GeV $\leq p_T\leq$ 15\,GeV and 
$E_T^{jet}>$ 5\,GeV.}
\label{muMhjet} 
\end{table}

This investigation is very CPU time consuming. Therefore, for a study of the 
optimum of $d\sigma/d\eta^h$ in various bins in $\eta^h$, we used the following
simplified approach. We fix the scale $M_F$ to the optimal value of 
Table~\ref{muMhjet},
$M_F = 0.5\,p_T$, and look for optima of $d\sigma/d\eta^h$ 
in the $(C_{\mu},C_M)$--plane for the $\eta^h$ bins 
of Fig.~\ref{figrapsca}.
The corresponding cross section is shown in Fig.~\ref{figrapsca} and we 
see that it is close
to the cross section obtained with $\mu=M=M_F=p_T/2$. It is encouraging to note
that the data of Fig.~\ref{fig1} are in good agreement with theory calculated with 
$C=0.5$ in the range $p_T > 7$\,GeV.


\begin{figure}[htb]
\begin{center}
\mbox{\epsfig{file=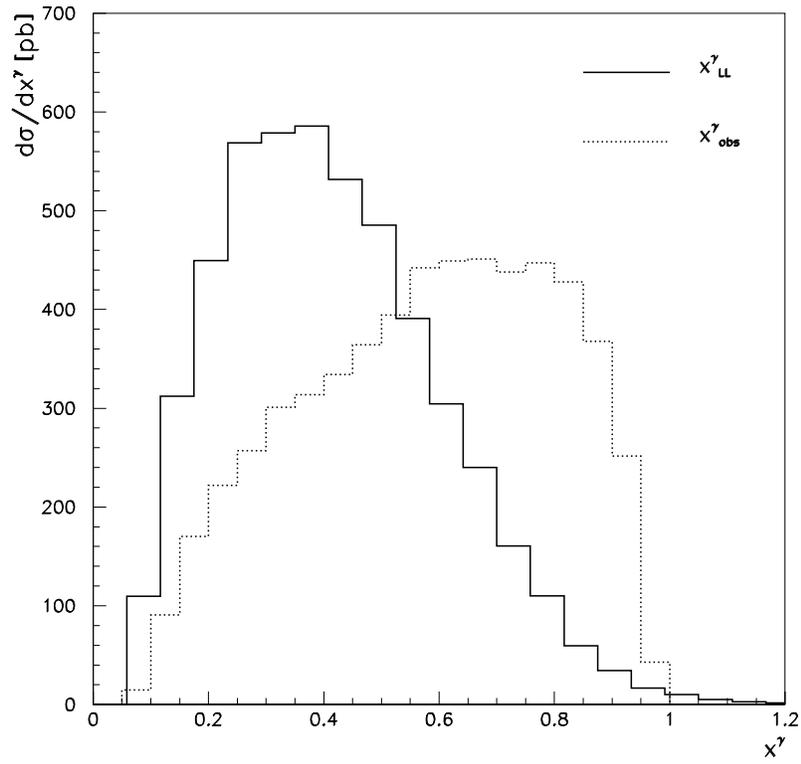,height=11.5cm}}
\end{center}
\caption{Comparison of $d\sigma/dx_{obs}^{\gamma}$ and 
$d\sigma/dx_{LL}^{\gamma}$, integrated over 
7\,GeV$\leq p_T\leq $15\,GeV, $E_T^{jet}\geq$ 5\,GeV, $|\eta^{jet,h}|<2$.}
\label{figxga}
\end{figure}

Let us now discuss the cross sections $d\sigma/dx_{obs}$ and  
$d\sigma/dx_{LL}$. Features of this cross sections are displayed in
Figs.~\ref{figxga} to~\ref{figlu0p}. 
Fig.~\ref{figxga} gives a clear illustration of the differences between 
the variables $d\sigma/dx_{obs}$ and  $d\sigma/dx_{LL}$. 
In the partonic variable $x^{\gamma}_{parton}$ (see eq.\,(\ref{xtrue})), 
the cross section would have a peak at $x^{\gamma}_{parton}\sim 1$, 
mainly due to the direct Born contribution which is proportional to 
$\delta(1-x^{\gamma}_{parton})$. This peak is shifted to lower values 
of $x^{\gamma}_{LL}$ and $x^{\gamma}_{obs}$ due to the relation 
$p_T^{h}=z\,p_T^{a}$, where $a$ is the "parent" parton of the hadron $h$. 
This shift is larger for the variable $x^{\gamma}_{LL}$, as 
discussed at the beginning of this section. 
We observe a similar, but much less pronounced pattern for the proton  
variable $x^{\rm{p}}$\,; 
the partonic distributions in the proton
decrease rapidly with $x^{\rm{p}}_{parton}$ and this behaviour is reflected in
the distributions of Fig.~\ref{figxp}, 
$d\sigma/dx^{\rm{p}}_{LL}$ having a smaller width 
than $d\sigma/dx^{\rm{p}}_{obs}$. 

\clearpage

\begin{figure}[htb]
\begin{center}
\mbox{\epsfig{file=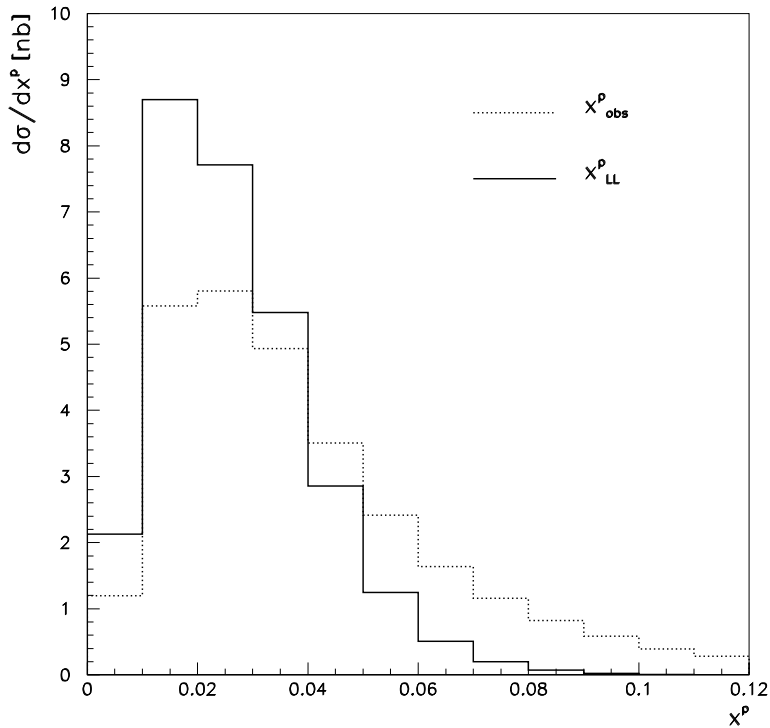,height=11cm}}
\end{center}
\caption{Comparison of $d\sigma/dx_{obs}^{\rm{p}}$ and
$d\sigma/dx_{LL}^{\rm{p}}$, integrated over 
7\,GeV$\leq p_T\leq $15\,GeV, $E_T^{jet}\geq$ 5\,GeV, $|\eta^{jet,h}|<2$.}
\label{figxp}
\end{figure}

Fig.~\ref{figxga3raps} displays a detailed study of the resolved and 
direct contributions as functions of $x^{\gamma}_{LL}$ and $x^{\gamma}_{obs}$. 
As already remarked, the shift of the peak to lower values of $x^{\gamma}$ 
is more pronounced for $x^{\gamma}_{LL}$ than for $x^{\gamma}_{obs}$.
The effect of cuts in $\eta^h$ and $\eta^{jet}$ are also clearly visible, 
the kinematic region $0\leq\eta^{h},\eta^{jet}\leq 2$ contributing to the 
low $x^{\gamma}$ domain, and the region  $-2\leq\eta^{h},\eta^{jet}\leq 0$ 
to the large $x^{\gamma}$ domain. 
The relative size of the resolved and
direct contributions as given in Fig.~\ref{figxga3raps} is of course 
dependent on
the scale choice ($\mu=M=M_F=p_T$ for Fig.~\ref{figxga3raps}). 
As already said at the beginning of this section, only the total cross section 
is a physical observable. The variation of the size of the direct and resolved
contributions with the scales varied by the factor $C$ 
is shown in Fig.~\ref{figxresdir}. 
For instance, we observe that for $0.5\leq C\leq 1$ the resolved component is
almost stable whereas the direct component strongly varies.

\begin{figure}[htb]
\begin{center}
\mbox{\epsfig{file=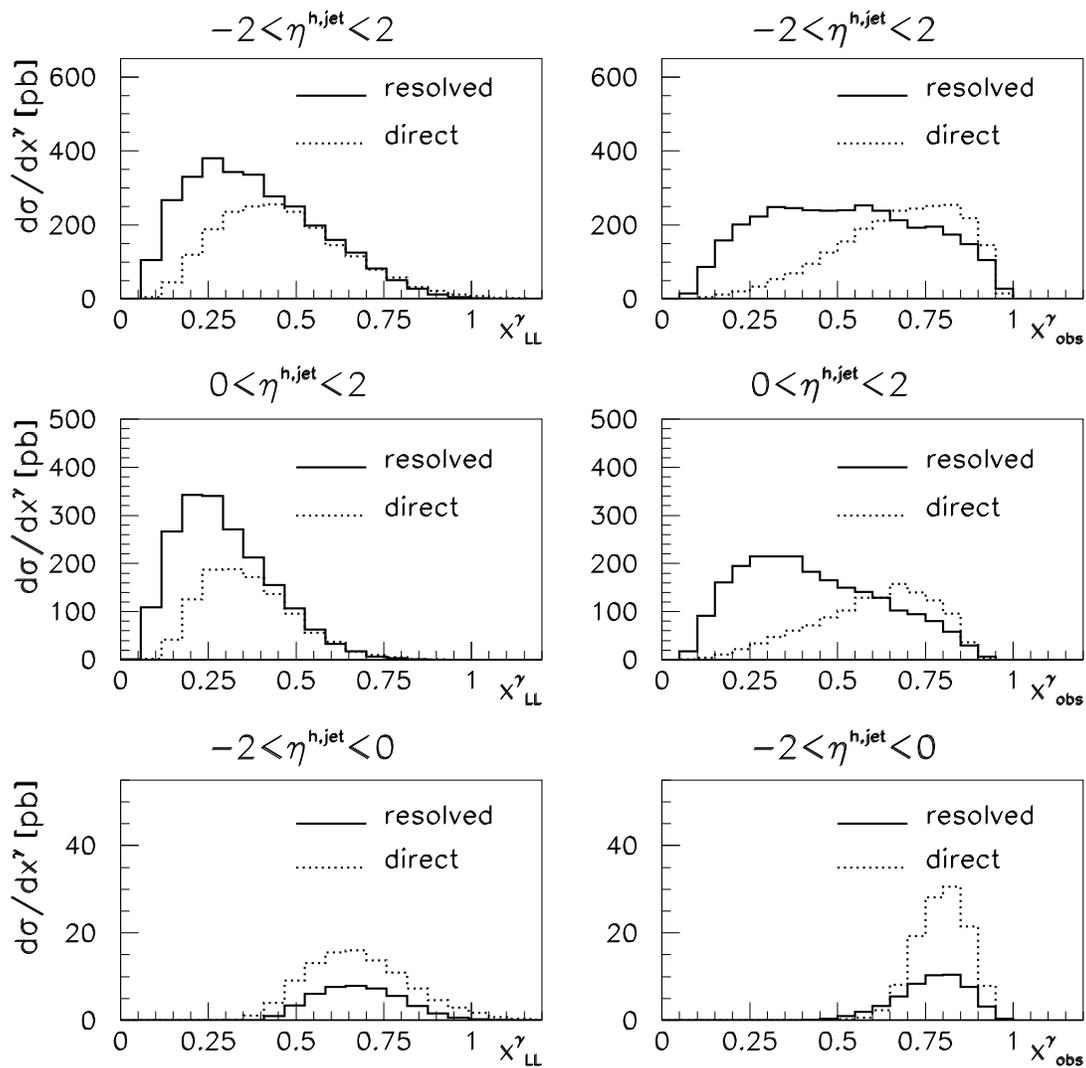,height=16cm}}
\end{center}
\caption{Comparison of resolved and direct contributions to 
$d\sigma/dx_{obs}^{\gamma}$ and $d\sigma/dx_{LL}^{\gamma}$
in different rapidity ranges.}
\label{figxga3raps}
\end{figure}

\begin{figure}[htb]
\begin{center}
\mbox{\epsfig{file=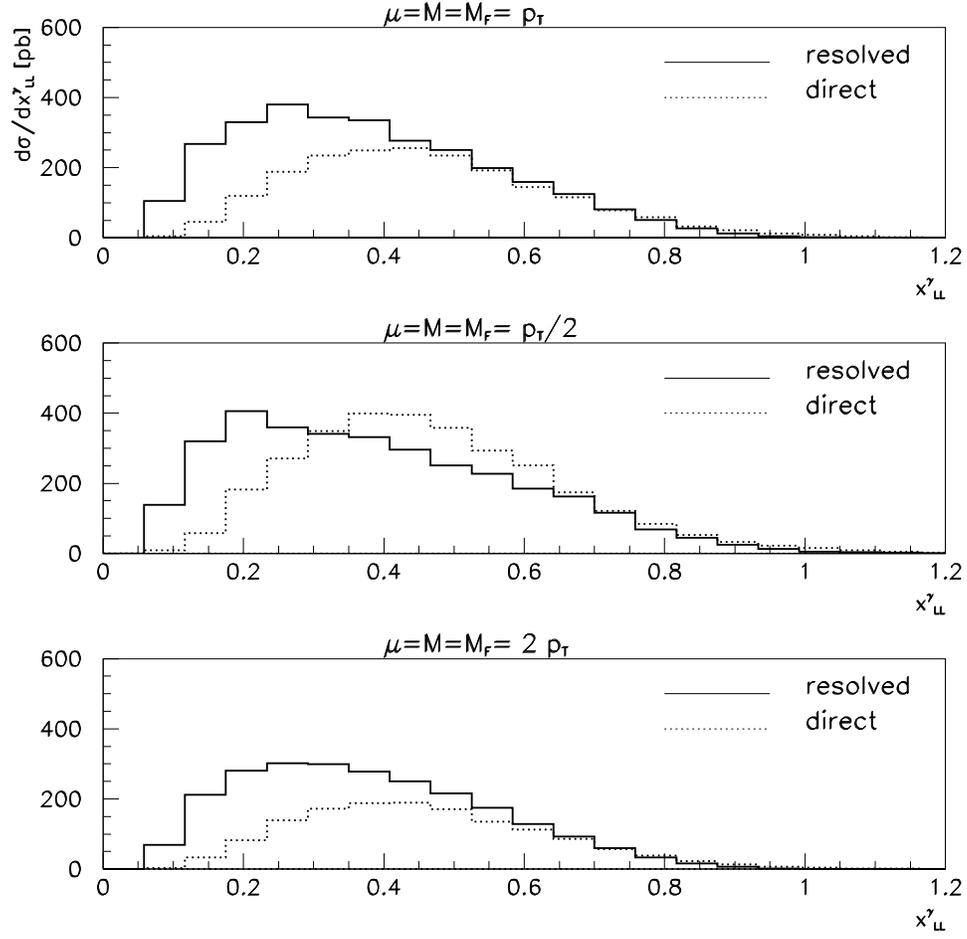,height=14cm}}
\end{center}
\caption{Comparison of resolved and direct parts for different scale choices,
integrated over the rapidity range $-2\leq\eta^{h,jet}\leq 2$.}
\label{figxresdir}
\end{figure}

\clearpage

The hadron-jet cross section offers the possibility
to measure the parton distributions in the proton and in the photon. 
The quark distributions are constrained by DIS experiments
(very well in the proton case, and with rather large errors 
in the photon case). Therefore, concerning the quark distributions,
the hadron-jet cross section  can only put some additional constraints 
on $F_{a/\gamma}(x,M)$. 
\begin{figure}[htb]
\begin{center}
\mbox{\epsfig{file=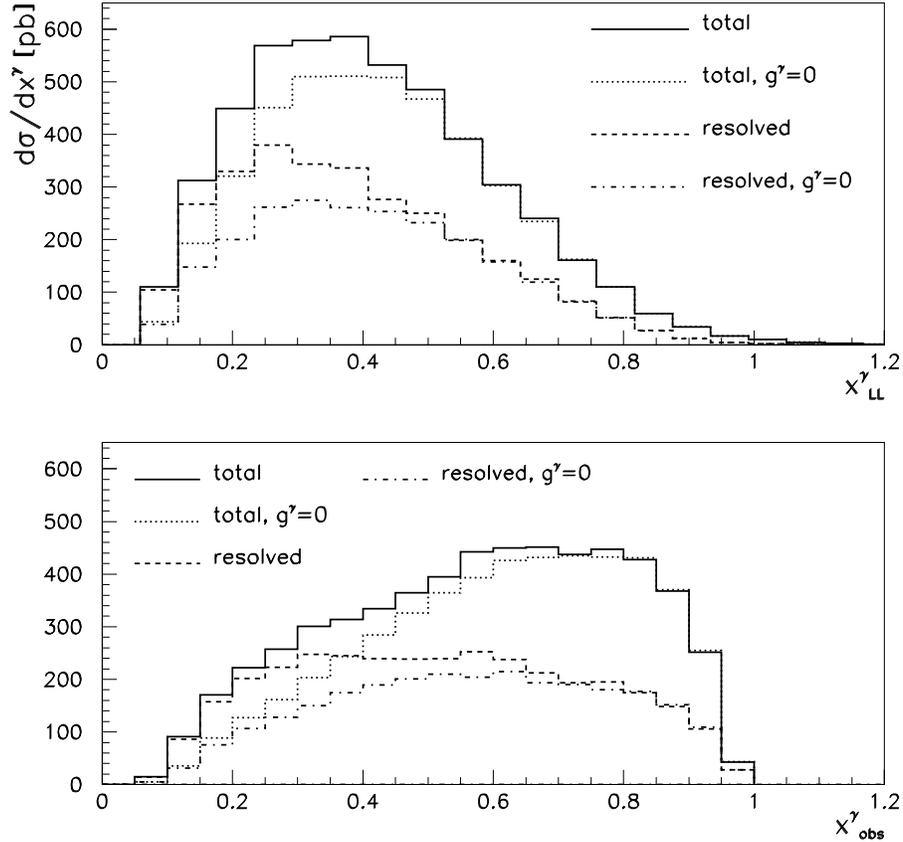,height=13cm}}
\end{center}
\caption{Contribution of the gluon in the photon to
$d\sigma/dx_{obs}^{\gamma}$ and
$d\sigma/dx_{LL}^{\gamma}$. The rapidities are integrated over the 
range $-2\leq\eta^{h,jet}\leq 2$.}
\label{figlu0ga}
\end{figure}
The situation with respect to the gluon distribution is different. 
The gluon distributions are not well determined in DIS experiments
because the virtual photon couples at leading order only to quarks,
such that the gluon distributions appear only at the level of higher order
corrections. This is not the case for the hadron-jet cross section 
where the gluon distributions appear already at the Born level. 
Therefore we expect an important sensitivity of the cross section to the gluon 
distributions, especially in the kinematic regions where $x_{parton}$
is small. This point is illustrated in Figs.~\ref{figlu0ga} to
 \ref{figlu0p}.
 
\begin{figure}[htb]
\begin{center}
\mbox{\epsfig{file=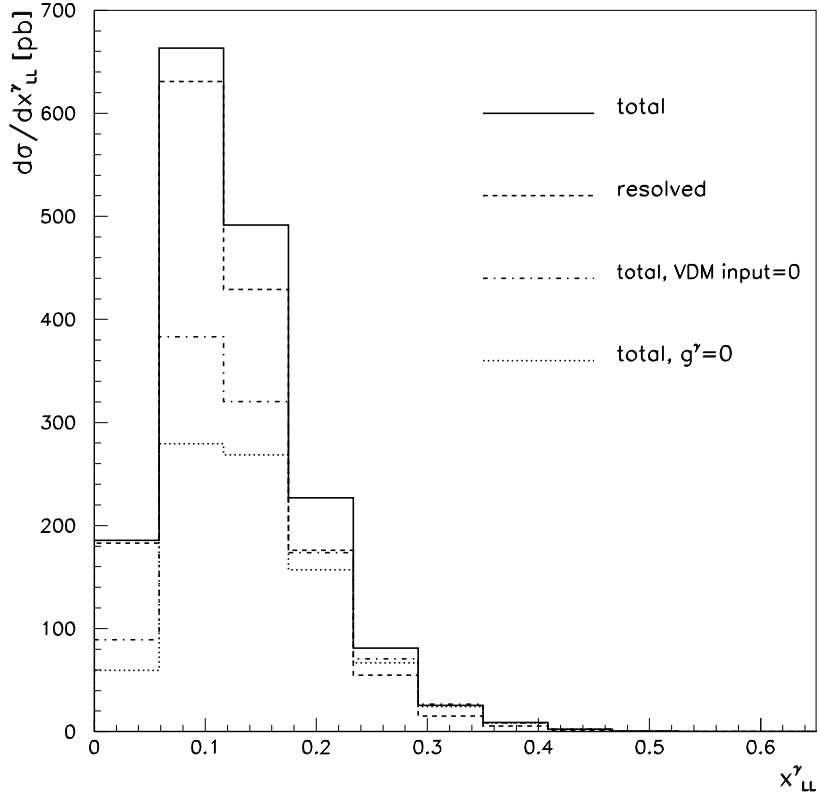,height=12cm}}
\end{center}
\caption{$d\sigma/dx_{LL}^{\gamma}$ with the hadron and jet rapidity cuts 
$1\leq \eta^h,\eta^{jet}\leq 3$.}
\label{figrap13}
\end{figure} 
 
In Fig.~\ref{figlu0ga}, we observe that the gluon in the photon makes a
large contribution at small $x^{\gamma}$, $x^{\gamma}\,\lsim \,0.25$. 
However, the cross section is small in this $x^{\gamma}$ region 
because the cuts $\eta^h\leq 2$ and $\eta^{jet}\leq 2$ forbid to reach 
low values of $x^{\gamma}$. Using different cuts which reinforce the small 
$x^{\gamma}$ region, namely $1\leq \eta^h,\eta^{jet}\leq 3$, we obtain
the results shown in Fig.~\ref{figrap13}. 
We see that the cross section in the small  $x^{\gamma}$ region is much 
larger and made up almost entirely by the resolved contribution. 
The AFG02 parton distributions for the photon allow to modify the normalisation 
of the non-perturbative VDM component. In order to exhibit the 
sensitivity to this component, we show a curve where the coefficient 
of this VDM input has been set to zero. We also show the magnitude of the 
gluon contribution to $d\sigma/dx_{LL}^{\gamma}$ in this kinematic range.  
We conclude from Fig.~\ref{figrap13} that the rapidity cuts 
$1\leq \eta^h,\eta^{jet}\leq 3$ 
select a kinematic region where the sensitivity 
of the hadron-jet cross section to the gluon distribution in the photon is 
very large. 


The contribution of the gluon in the proton turns out to be large even 
with rapidities integrated in the whole range $-2\leq\eta^{h,jet}\leq 2$, 
as can be seen from Fig.~\ref{figlu0p}. We also observe that 
the effect of using different parton distributions for the proton, namely 
the MRST99~\cite{mrst99}, the new MRST01~\cite{mrst01} and the 
CTEQ6M~\cite{cteq6} sets, is quite large  and mainly due to the different 
shape of the gluon in the different sets. 
Although the variation of the cross section when varying all scales simultaneously
between $0.3\leq C\leq 2$ is larger than the variation due to the different pdf
sets, the scale variations rather produce an overall shift of the curve, but do
not change the shape. On the other hand,
the fact that the gluon distribution 
of MRST01 peaks at higher $x$ values than the one of CTEQ6M 
(see Fig.\,16 of \cite{cteq6}) is clearly reflected in Fig.\,\ref{figlu0p}.
We also note that the region $x^{\rm p}_{LL}\approx 0.02$, 
corresponding to $x^{\rm p}_{parton}\approx 0.05$, is an interesting $x$-range 
because it lies in the window between constraints from HERA for lower $x$
and the Tevatron jet data for higher $x$. 
Therefore the cross section 
$d\sigma/dx_{LL}^{\rm{p}}$ can serve to further constrain 
 the parton distributions of the proton.  

\begin{figure}[htb]
\begin{center}
\mbox{\epsfig{file=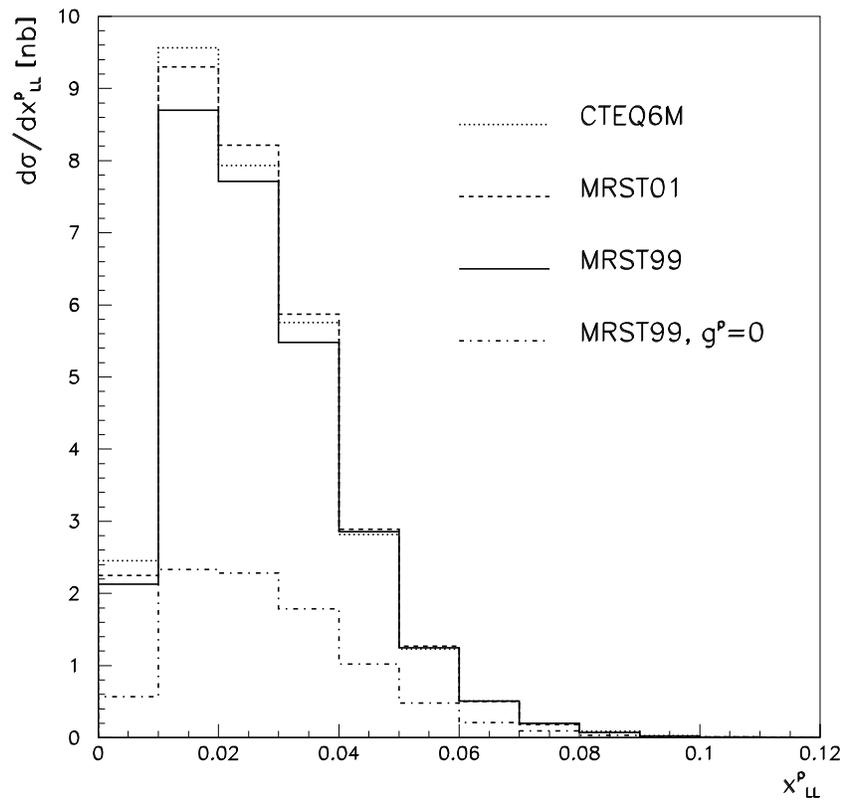,height=12cm}}
\end{center}
\caption{$d\sigma/dx_{LL}^{\rm{p}}$ calculated with different pdf sets and
size of the gluon in the proton\,;
rapidities integrated over the range $-2\leq\eta^{h,jet}\leq 2$.}
\label{figlu0p}
\end{figure}

\clearpage

\section{Conclusion}
We have studied the photoproduction of inclusive large-$p_T$ charged hadrons 
and the production of a charged hadron plus a jet. 
For the inclusive case we compared the $p_T$- and rapidity 
distributions to H1 data~\cite{18r} and found 
reasonable agreement. However, for a value of $p_T^{\rm{min}}$ as low as 
3\,GeV, the dependence of the NLO result on the renormalisation 
and factorisation scales is very large. Only for $p_T^{\rm{min}}\,\gsim 7$\,GeV
a plateau where the cross section is approximately stable against
scale variations could be found. 
We also studied the effect of using different fragmentation function 
parametrisations~\cite{7r,8r,9r} and compared to a previous analysis of 
Kniehl, Kramer and P\"otter~\cite{8r}. 
For the parton distributions in the photon, we used the new AFG02 
parametrisations~\cite{12r}. 

For the hadron-jet cross section, we studied the rapidity distributions 
and the cross sections $d\sigma/dx^{\rm{p}},d\sigma/dx^{\gamma}$. 
We analysed the difference between the partonic momentum fractions 
$x^{\rm{p},\gamma}_{parton}$ and the observables  $x^{\rm{p},\gamma}_{obs}$
defined via the observed transverse momenta and rapidities of the 
hadron and the jet. We further proposed a variable  $x^{\rm{p},\gamma}_{LL}$
which does not require the measurement of the jet transverse energy. 

We also carried out an exhaustive study of the scale dependence.  
We found a stability region for the cross section integrated over 
7\,GeV $\leq p_T^h\leq$ 15\,GeV, $E_T^{jet}>$ 5\,GeV and 
performed a scale optimisation.
 
Finally, we investigated the possibility to constrain the parton 
distributions (in particular the gluon distributions) 
in the photon and in the proton via the hadron-jet cross 
section. We show how rapidity cuts can increase the sensitivity to 
the gluon distributions in the photon. 
We also found a rather large sensitivity to the parton distributions 
in the proton. We show a comparison of the MRST99~\cite{mrst99} 
distributions to the new 
MRST01~\cite{mrst01} and CTEQ6M~\cite{cteq6} distributions.

\hspace*{\parindent}

\noindent{\bf \large Acknowledgements}

G.H. would like to thank the LPT Orsay for its kind hospitality in May 02.
This work was supported by the EU Fourth Training Programme  
''Training and Mobility of Researchers'', network ''Quantum Chromodynamics
and the Deep Structure of Elementary Particles'',
contract FMRX--CT98--0194 (DG 12 - MIHT).

\end{document}